\documentstyle[aps,twocolumn,epsf]{revtex}

\def\be{\begin{equation}}
\def\ee{\end{equation}}

\newcounter{fig}

\catcode`\@=11   

\newcommand{\fcaption}[1]{\vspace{1ex}   
        \refstepcounter{figure}   
        \setbox\@tempboxa = \hbox{\footnotesize {\bf Fig.~\thefigure.} #1}   
        \ifdim \wd\@tempboxa > 7cm   
           {\begin{center}   
        \parbox{7cm}{\footnotesize\baselineskip=8pt {\bf Fig.~\thefigure.} #1}   
            \end{center}}   
        \else   
             {\begin{center}   
             {\footnotesize {\bf Fig.~\thefigure.} #1}   
              \end{center}}   
        \fi}

\begin{document}

\title{The Atomic Slide Puzzle: Self-Diffusion of an Impure Atom.
}

\author{ O.B{\'e}nichou$^{1}$
 and G.Oshanin$^{2}$
}

\address{ $^{1}$ Laboratoire de Physique de la Mati{\`e}re Condens{\'e}e,
Coll{\`e}ge de France, 11 place M.Berthelot, 75231 Paris Cedex 05, France
} 

\address{  $^{2}$ Laboratoire de Physique Th{\'e}orique des Liquides (CNRS - UMR 7600), 
Universit{\'e} Pierre et Marie Curie,
4 place Jussieu, 75252 Paris Cedex 05, France
}

\address{\rm (Received: )}
\address{\mbox{ }}
\address{\parbox{14cm}{\rm \mbox{ }\mbox{ }
In a series of recent papers 
\cite{saar1,saar2,saar3} van Gastel et al 
have presented first experimental evidence, based on 
 a series of STM images, 
 that impure, 
Indium atoms, embedded 
 into the first, close-packed layer of a ${\rm Cu(001)}$ surface, are not localized 
but
make concerted, 
long excursions. 
Such excursions 
occur due
to continuous reshuffling of 
the surface 
following the position exchanges of both impure and host ${\rm Cu}$ 
atoms with the naturally occuring surface vacancies. 
Van Gastel et al have also formulated an original 
lattice-gas type model with asymmetric exchange probabilities, 
whose
numerical solution is in a good agreement 
with the experimental data. In this paper we propose
an exact lattice solution of several versions of this model.
}}
\address{\mbox{ }}
\address{\parbox{14cm}{\rm PACS No:  68.35.Fx, 05.40.Fb, 66.30.Lw, 07.79.Cz }}
\maketitle

\makeatletter
\global\@specialpagefalse

\makeatother

\pagebreak

Surface mobility is usually believed to be localized in the vicinity of the terrace
steps and kinks of crystal surfaces. These sites form the natural locations for atoms
to attach to or to
detach from the terraces; the diffusion of adatoms along or between the steps
is known to induce their roughening  
and is responsible for the two-dimensional diffusive motion of
adatom and vacancy islands \cite{1,2,3,31}. 
By contrast, the atoms within the close-packed terraces that are not in the immediate
vicinity of steps have often been  considered as immobile since they
are tightly packed by their neighbors. Although some suggestions have been made that
the surface vacancies may diffusive themselves within the surface layers (see, e.g. Ref.\cite{4}),  
up to a very recent time there were no 
experimental techniques allowing for the direct observation of 
the 
 vacancies' or atoms' diffusion  
within the close-packed surfaces.

In a recent Letter and two accompanying papers
van Gastel et al \cite{saar1,saar2,saar3} presented an 
indirect experimental evidence that, remarkably, 
 the atoms of the close-packed
surfaces do undergo themselves 
continuous 
random motion. 
In a set of ingenious STM 
experiments, van
Gastel et al \cite{saar1,saar2,saar3} have managed to
introduce impure (Indium) atoms within the 
first layer of a Cu(001) surface 
and were able to follow their positions
by 
analyzing the series of consequtive
STM images. 
They found  
that the embedded ${\rm In}$
atoms make concerted, long excursions 
within the first surface layer, which motion can be only explained (see Refs.\cite{saar1}
and \cite{saar2,saar3}
for more details) by continuous reshuffling of 
the surface in a way which resembles a slide puzzle
due to position exchanges of both the ${\rm Cu}$ and ${\rm In}$ 
atoms with the naturally occuring surface vacancies \cite{estimate}.

Van
Gastel et al \cite{saar1,saar2}
also proposed a simple 
model describing the ${\rm In}$ atom
dynamics. 
In this model, the authors considered 
a terrace on a stepped ${\rm Cu}$ surface as a  
finite square lattice all $L \times L$ 
sites of which
except two are filled with 
${\rm Cu}$ atoms; the  ${\rm In}$ atom is initially placed at the lattice origin and
the vacancy - at one of the adjacent sites. 
Both ${\rm In}$ and ${\rm Cu}$
atoms move randomly along the lattice by exchaning their positions with the vacancy,
whose random walk terminates (and may reappear again) as soon as it reaches
the lattice boundary.

A salient feature of the physical situation is that, 
due to 
the difference of the ${\rm Cu-Cu}$ and the
${\rm Cu-In}$ interactions, the vacancy being at the 
adjacent to the ${\rm In}$ atom 
site has 
a preference to exchange its position with 
the ${\rm In}$ atom, compared to three adjacent 
${\rm Cu}$ atoms.   
The chemical specificity of atoms is then
taken into account by introducing unequal
hopping probabilities; that is, in case when one of atoms adjacent to the vacancy is
the ${\rm In}$ atom, the vacancy exchanges its position with the latter 
or with one of three ${\rm Cu}$ atoms with the probabilities
\begin{equation}
\label{1}
p_{\rm In} = \frac{1}{4}(1 + \epsilon), \;\;\;
p_{\rm Cu} = \frac{1}{4}(1 - \frac{\epsilon}{3}),
\end{equation}
respectively. For the chemical species under study
and  at the room temperature the asymmetry parameter
$\epsilon$   
assumes the value  $\epsilon^* = 3 - 10^{-6}$, 
according to the 
Embedded Atom Model calculations \cite{saar1}. 
In case when all four adjacent to the vacancy atoms are the  ${\rm Cu}$ atoms,
the vacancy exchanges its position with any of them with the probability $= 1/4$. 
As a matter of fact, numerical simulations \cite{saar1} show that the difference in the exchange probabilities
has a significant impact on the ${\rm In}$ atom displacements: the average length of excursions
 is $2.2$ times larger
than in the $\epsilon = 0$ case.
 Monte Carlo simulations of this model produce the results which
are in a good agreement with the experimentally obtained jump length distribution 
\cite{saar2}
and, hence, confirm the vacancy-assisted mechanism of the ${\rm In}$ atom dynamics
proposed in Ref.\cite{saar1}. 

Exact solution of this model is known only in the limit
$\epsilon = 0$ \cite{hilhorst1} (see also Ref.\cite{henk} 
for the analysis in case of arbitrary vacancy concentrations $\rho$).  
Van Gastel et al  \cite{saar1,saar2} have, however,  
furnished  
an approximate,  
continuous-space solution for  the jump length distribution function 
supposing that the boundary conditions 
associated with the vacancy creation/annihilation appearing at
the terrace edges can be modeled 
by introducing some, 
$\em a \; priori$ 
given vacancy life-time. 
Despite the fact that it is possible to 
find such values 
of this heuristic
parameter at which 
the
obtained results for the jump length probability distribution 
agree quite well with the experimental ones,  
it is, however, difficult to claim that such 
a comparison can 
provide meaningful values of 
other physical parameters. 
Clearly, an 
assumption that the vacancy can spontaneously disappear at any lattice point, 
not necessarily at the terrace edges, might be a useful simplification but 
is not controllable. On the other hand,  
it is well-known that in two-dimensions the continuous-space description may 
incur significant errors in the numerical factors, which
will result in modified values of the physical parameters. 
Lastly,  the continuous-space approach is hardly appropriate  
for studying the probability
distributions on scales of order of several
lattice spacings only. 
Consequently, for meaningful interpretation of 
the experimental data an exact lattice solution of the model is highly
desirable.

In this paper  we present an exact lattice solution 
of several versions of the model
introduced by  van
Gastel et al \cite{saar1,saar2,saar3} for arbitrary values of the parameter
$\epsilon$, $-1 \leq \epsilon < 3$, 
where positive (negative) values of $\epsilon$  
correspond to the physical situation in 
which the ${\rm In}$ atom has a preference (reduced probability) for
the position exchanges with the vacancy.     
More specifically, we consider a situation with
a single vacancy performing the just described random walk on a finite lattice with
periodic boundary conditions, which mimic annihilation/creation of the vacancy at the
terrace boundaries and also solve the case with a single vacancy and
 $L  = \infty$, which is
appropriate for evaluation of 
the transient dynamics. We note that in view of a very low concentration $\rho$ 
of the vacancies \cite{estimate}, this transient
regime may be very important and may 
persist over a wide time range.
In both situations we determine exactly the
leading long-time asymptotical behavior of the mean-square displacement (MSD) and the 
diffusivity of a single ${\rm
In}$ atom, as well as of the probability that the ${\rm
In}$ atom appears at position ${\bf X}$ at time moment $t$. All our results will be presented in 
dimensionless form as 
functions of the discrete time $n$ and 
for the
lattice spacing set equal to unity. 
Dimension-dependent forms can be  obtained by
rescaling $n \to t/\Delta t$ and  
${\bf X} \to {\bf X}/\sigma$, where $\Delta t$ $( \approx 10^{-8} s)$ is the typical time interval between successive 
jumps
and $\sigma$ $( \approx 2.5$ $\AA$) is the intersite distance \cite{saar1}.  

We begin with more precise definition of the model. 
Consider a two-dimensional, periodic in both  $x_1$ and $x_2$ directions, 
square lattice of unit spacing 
all $L \times L$ sites of which except two (a
vacancy and an ${\rm In}$ atom)
are filled by ${\rm Cu}$ atoms (see Fig.1). 
The ${\rm In}$ atom is placed initially at the origin, while the vacancy is at
position with the vector ${\bf Y} = {\bf e_{-1}}$. 
Here, ${\bf e}_{  \nu}$, $\nu\in\{\pm1,\pm2\}$,  
denotes  unit lattice vectors;   ${\bf
e_{1}}$ (${\bf
e_{-1}}$) is the unit vector in the positive (negative) $x_1$-direction, while ${\bf
e_{2}}$ (${\bf e_{- 2}}$) is the unit vector in the positive (negative) 
$x_2$-direction. 

Next, at each tick of the clock, $n = 1, 2, 3, \ldots$,  a vacancy 
exchanges its position with one of four neighboring atoms according to the
probabilities defined in Eqs.(\ref{1}). 
That is, it has a preference (or, on contrary, a reduced probability 
for $\epsilon < 0$) to
exchange its position with the  ${\rm In}$ atom, when this one is 
at the neighboring site, and no preferency for exchange with any of atoms when all
four adjacent to the vacancy sites are occupied by the ${\rm Cu}$ atoms. Hence, the
vacancy moves randomly displacing the atoms
in its path, including the ${\rm In}$ atom; apart from four "defective" sites adjacent
to the ${\rm In}$ atom, it performs conventional symmetric random walk.

\begin{figure} \begin{center}   
  \fbox{\epsfysize=6cm\epsfbox{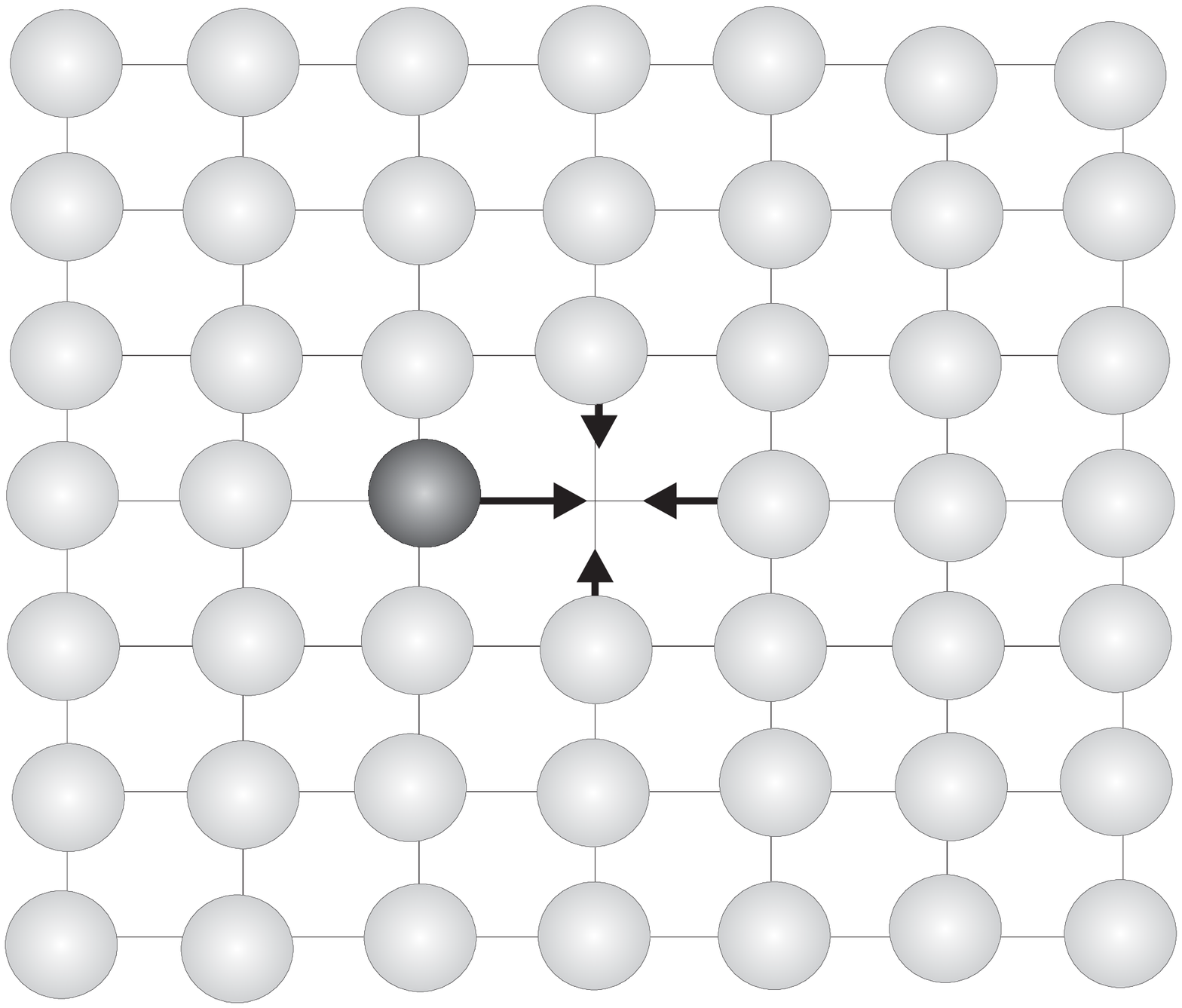}}   
   \fcaption{An illustration of the vacancy-assisted diffusion mechanism of the ${\rm
In}$ atom (black sphere) on a lattice filled with the ${\rm Cu}$ atoms (grey
spheres). The arrows depict the asymmetric hopping probabilities of the  ${\rm
In}$ and  ${\rm Cu}$ atoms.}   
  \label{tra}   
\end{center}\end{figure}    

Let
$P_n({\bf X})$ be the probability that the
 ${\rm
In}$ atom, which starts its random walk at the origin, appears 
at the site ${\bf X}$ at time moment $n$, given that 
the vacancy is initially at the site ${\bf e_{-1}}$. 
Then, following the approach of Ref.\cite{hilhorst1}, we 
write down 
the equation obeyed by 
$P_n({\bf X})$ as:
\begin{eqnarray}
&&P_n({\bf X})= 
\sum_{p=1}^{+\infty}\sum_{m_1=1}^{+\infty}\ldots
\sum_{m_p=1}^{+\infty}
\sum_{m_{p+1}=0}^{+\infty}\sum_{\nu_1}\ldots\sum_{\nu_p} \times\nonumber\\
&\times& \left(1-\sum_{j=0}^{m_{p+1}}F_j({\bf
0}|-{\bf e}_{{  \nu}_{\bf p}})\right) \; F_{m_1}({\bf 0}|{\bf
e}_{{  \nu}_{\bf 1}}|{\bf e_{-1}}) \times  \nonumber\\
&\times& \left(\prod_{i = 2}^{p} F_{m_i}({\bf 0}|{\bf e}_{{  \nu}_{\bf i}}|-{\bf
e}_{{  \nu}_{\bf i-1}}) \right)  \delta_{m_1+\ldots+m_{p+1},n} \times  \nonumber\\
&\times& \delta_{{\bf
e}_{{  \nu}_1}+\ldots+{\bf
e}_{{  \nu}_p},{\bf X}} +   \delta_{{\bf X},{\bf 0}}\left(1-\sum_{j=0}^n F_j({\bf
0}|{\bf e_{-1}})\right),
\label{Ptr}
\end{eqnarray}
where
$F_n({\bf 0}|{\bf e_\mu})$ is the probability that the
vacancy, which starts its random walk at the site ${\bf e_\mu}$,
 arrives at the origin  ${\bf 0}$ for the first time at the $n$-th step, while
$F_n({\bf 0}|{\bf e}_{  \nu}|{\bf e_\mu})$ stands for the
$conditional$ probability that the vacancy, 
which starts its random walk at the site ${\bf e_\mu}$, 
 appears  at the origin for the first
time at the $n$-th step, being at the $(n - 1)$-th step at the site 
${\bf e}_{  \nu}$.

General solution of Eq.(\ref{Ptr}) can be written down in form of the following 
contour integral (see, Refs.\cite{hilhorst1} and \cite{olivier}): 
\begin{eqnarray}
\label{4}
P_n({\bf X})=\frac{-i}{(2 \pi)^3}\oint_{\cal C} \frac{{\rm
d}\xi}{\xi^{n+1}} \int {\rm d}{\bf k} \;
e^{i ({\bf k} \cdot {\bf X})} \; P_{\xi}({\bf k}),
\end{eqnarray}
where the contour of integration ${\cal C}$ encircles the origin counterclockwise,
$({\bf k} \cdot {\bf X})$ denotes the scalar product, 
and
\begin{equation}
\label{P}
P_{\xi}({\bf k})=\frac{1}{1-\xi}\left(1+{\cal D}^{-1}_{\xi}({\bf
k})\sum_{\mu}U^{(\mu)}_{\xi}({\bf k}) F_{\xi}({\bf
0}|{\bf e}_{  \mu}|{\bf e_{-1}})\right)
\end{equation}
In Eq.(\ref{P})
${\cal D}_{\xi}({\bf k})$ denotes  
the following determinant:
\begin{equation}
\label{D}
{\cal D}_{\xi}({\bf k})\equiv{\rm det}({\bf  I-T_{\xi}}({\bf k})),
\end{equation} 
the 
 elements $\Big({\rm {\bf T}}_{\xi}({\bf k})\Big)_{\nu,\mu}$ of the $4 \times 4$
matrix ${\rm {\bf T}}_{\xi}({\bf k})$ obey:
\begin{equation}
\Big({\rm {\bf T}}_{\xi}({\bf k})\Big)_{\nu,\mu} \equiv \exp\Big(i ({\bf k} \cdot {\bf
e_{  \nu}})\Big) \; F_{\xi}({\bf 0}|{\bf e}_{\nu}|{\bf e}_{-\mu}),
\end{equation}
the indices $\nu$ and $\mu$ assume successively the values $1,-1,2,-2$;  
$ F_{\xi}({\bf 0}|{\bf e}_{\nu}|{\bf e}_{-\mu})$ is the generating function of the
conditional first-visit probability $F_{n}({\bf 0}|{\bf e}_{\nu}|{\bf e}_{-\mu})$, i.e.
\begin{equation}
 F_{\xi}({\bf 0}|{\bf e}_{\nu}|{\bf e}_{-\mu}) \equiv \sum_{n =0}^{\infty} 
F_{n}({\bf 0}|{\bf e}_{\nu}|{\bf e}_{-\mu}) \; \xi^n,
\end{equation}
and the matrix $U^{(\mu)}_{\xi}({\bf k})$ is given by
\begin{eqnarray}
\frac{U^{(\mu)}_{\xi}({\bf k})}{{\cal D}_{\xi}({\bf k})}\equiv   e^{i ({\bf
k} \cdot {\bf e}_{  \mu})} \; \sum_{ \nu} 
(1-e^{-i ({\bf k} \cdot {\bf e}_{  \nu})})({\rm I} -{\rm T}_{\xi}({\bf
k}))^{-1}_{\nu,\mu}.
\end{eqnarray}
In turn, the MSD of the  ${\rm In}$ atom is defined as
\begin{equation}
\label{X}
\overline{{\bf X}^2} \equiv 4 D_n(\epsilon) n \equiv \frac{i}{2 \pi} \oint_{\cal C} \frac{{\rm
d}\xi}{\xi^{n+1}}  \left. \frac{d^2 P_{\xi}({\bf k})}{d {\bf k}^2} \right|_{|{\bf k}| = 0},
\end{equation}
where $P_{\xi}({\bf k})$  is determined by Eq.(\ref{P}).

We note now that as far as we are interested only in the leading large-$n$ 
 behavior  of 
 $P_n({\bf X})$ and of the MSD, we may constrain 
ourselves to the study of the asymptotic behavior of 
the generating function
$\widetilde{P}^{(tr)}({\bf k};\xi)$ in the vicinity of 
its singular point nearest to
$\xi=0$. 
Omitting the details of straightforward, but rather tedious calculations, we find that 
 the leading, in the combined ${\bf k
}\to 0$ and $\xi \to 1^-$ limit,  behavior of  ${\cal D}_{\xi}({\bf k})$  follows
\begin{equation}
\label{11}
{\cal D}_{\xi}({\bf k}) = {\cal F}_1(\epsilon) {\bf k}^2 + {\cal F}_2(\epsilon) (1-\xi) + \ldots ,
\end{equation}
where
\begin{eqnarray}
&&{\cal F}_1(\epsilon) = - \frac{(\pi - 2) (\pi + 2 L^2) (\epsilon - 3)^2}{2 (5 \pi \epsilon - 16 \epsilon -
3 \pi)} \times \nonumber\\
&\times& \frac{(- 14 \epsilon L^2 +  6 \pi \epsilon L^2 - 7 \pi \epsilon + 6 \pi L^2 - 6 L^2 - 3 \pi)}{((8 \epsilon - 3
\pi \epsilon - 3 \pi) L^2  + 4 \pi \epsilon)^2},
\end{eqnarray}
and
\begin{eqnarray}
\label{f2}
{\cal F}_2(\epsilon) &=& \frac{2 (\pi - 2) (3 - \epsilon) ((3 - \epsilon) L^2 + 5 \epsilon - 3 )}{3 (1 +
\epsilon) (16 \epsilon - 5 \pi \epsilon + 3 \pi) } \times \nonumber\\
&\times& \frac{((14 \epsilon - 6 \pi \epsilon - 6 \pi + 6) L^2 + 7 \pi \epsilon + 3 \pi)^2}{((8 \epsilon - 3
\pi \epsilon - 3 \pi) L^2  + 4 \pi \epsilon)^2}
\end{eqnarray}

\begin{figure} \begin{center}   
  \fbox{\epsfysize=5.5cm\epsfbox{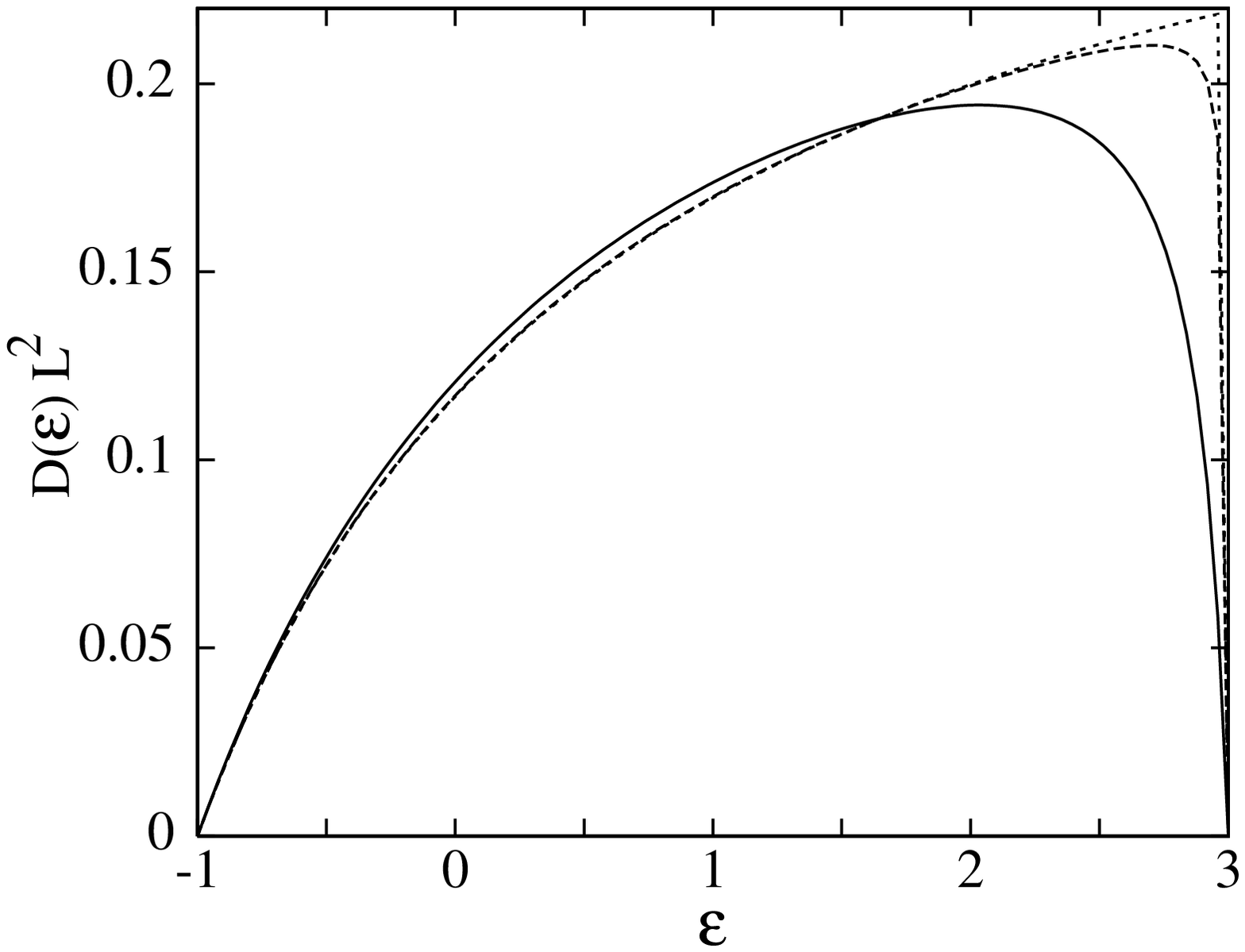}}   
   \fcaption{Plot of $D(\epsilon)$, 
Eq.(\ref{diff}), versus the parameter $\epsilon$ for different values of the
terrace width. The solid line corresponds to $L = 15$, the dashed - to $L = 40$ and the dotted - to $L = 600$,
respectively. For notational convenience $D(\epsilon)$ is multiplied by the terrace area $L^2$.}   
\label{tara}   
\end{center}\end{figure}

Then, substituting these expressions into Eqs.(\ref{P}) and (\ref{X}), and noticing that
\begin{equation}
\label{14}
\sum_{\mu}U^{(\mu)}_{\xi}({\bf k}) F_{\xi}({\bf
0}|{\bf e}_{  \mu}|{\bf e_{-1}}) = {\cal F}_1(\epsilon) \; {\bf k}^2 + \; \ldots \; , 
\end{equation}
we find that as $n \to \infty$, 
the diffusivity $D_n(\epsilon)$ of the ${\rm In}$ atom,  Eq.(\ref{X}), 
tends to a constant value 
\begin{eqnarray}
\label{diff}
&&D(\epsilon) = \frac{3 (1 + \epsilon) (3 - \epsilon)}{4 \Big((3 - \epsilon) L^2 + 5 \epsilon -3 \Big)}  \times \nonumber\\
&\times& \frac{(\pi + 2 L^2)}{\Big((6 \pi \epsilon - 14
\epsilon + 6 \pi - 6) L^2 - 7 \pi \epsilon - 3 \pi\Big)}.
\end{eqnarray}
Note that $D(\epsilon)$ 
appears to be a non-trivial, non-monotoneous 
function of the parameter $\epsilon$ (see, Fig.2);  $D(\epsilon)$
attains a maximal value when $\epsilon =
\epsilon_{c} \approx 3 - 12 \sqrt{\pi - 2}/L$, and is exactly equal to zero for $\epsilon =
-1$ and $\epsilon =
3$. In the case $\epsilon =
-1$, the ${\rm In}$ atom does not move at all, 
since it is not allowed to 
change its position with the vacancy, while
in the $\epsilon =
3$ case it gets localized 
because at each time step it is forced to exchange its position with the single
available vacancy. 
Note that such a "localization effect" for $\epsilon = 3$ 
is, of course, specific to the situation with a single available vacancy; 
as a matter of fact, the indium atom can get delocalized already if a 
second vacancy is present.  
Consequently, the result in Eq.(\ref{diff})
concerns only the leading at $\rho = 0$ behavior. 
For finite but 
small $\rho$, there should be a correction term  to Eq.(\ref{diff}) 
proportional to $\rho^2$. On the other hand, in view of a very small 
concentration of the naturally occuring vacancies \cite{estimate},
observation of such correction terms (non-zero values of $D(\epsilon=3)$) 
would require very 
large observation times. 
Note also that the ratio $D(\epsilon = \epsilon^*)/D(\epsilon = 0)$ is
approximately equal to $1.8$ for 
terrace width $L = 400$, which shows that, indeed, 
asymmetric exchange 
probabilities have also a strong effect on the 
 ${\rm In}$ atom diffusivity, not only on the average excursion length
\cite{saar1}.
   
Lastly, following Ref.\cite{hilhorst1}, we get from Eqs.(\ref{4}) and (\ref{P}), 
and the asymptotical expansions in Eqs.(\ref{11}) to (\ref{14}), that for sufficiently large $n$, such that
$L^2 \ln(L) \ll n \ll L^4$, the scaling variables
 $\xi_1 = x_1/D(\epsilon) n$ and $\xi_2 = x_2/D(\epsilon) n$ 
are distributed according to a Gaussian law:
\begin{equation}
\label{G}
P(\xi_1,\xi_2) = (2 \pi)^{-1} \; \exp\Big( - (\xi_1^2 + \xi^2_2)/2 \Big).
\end{equation}

We turn now to the case $L = \infty$, which will allow us to obtain a meaningful transient 
behavior. Again, omitting the details of  calculations, 
we find that the leading, in the combined ${\bf k
}\to 0$ and $\xi \to 1^-$ limit, behavior of  ${\cal D}_{\xi}({\bf k})$ obeys:
\begin{equation}
\label{N}
{\cal D}_{\xi}({\bf k}) = \tilde{{\cal F}}_1(\epsilon) \;  {\bf k}^2 - \tilde{{\cal F}}_2(\epsilon) 
\Big(\ln{(1-\xi)}\Big)^{-1} + \; \ldots \; ,
\end{equation}
where
\begin{equation}
\tilde{{\cal F}}_1(\epsilon) = - 2 \frac{(\pi - 2) (\epsilon - 3)^2 (3 \pi \epsilon - 7 \epsilon + 3 \pi - 3)}{(5 \pi \epsilon - 16 \epsilon -
3 \pi) (3 \pi \epsilon - 8 \epsilon + 3 \pi)^2},
\end{equation}
and
\begin{equation}
\label{K}
\tilde{{\cal F}}_2(\epsilon) = - \frac{8 \pi (\pi - 2) (\epsilon - 3)^2 (3 \pi \epsilon
 - 7 \epsilon + 3 \pi - 3)^2}{3 (1 + \epsilon) (5 \pi \epsilon - 16 \epsilon -
3 \pi) (3 \pi \epsilon - 8 \epsilon + 3 \pi)^2    },
\end{equation}
while the sum $\sum_{\mu}U^{(\mu)}_{\xi}({\bf k}) F_{\xi}({\bf
0}|{\bf e}_{  \mu}|{\bf e_{-1}})$ still obeys Eq.(\ref{14}) with ${\cal F}_1(\epsilon)$ replaced by $\tilde{{\cal
F}}_1(\epsilon)$.
Consequently, we find that in the leading in 
$n$ order the diffusivity of the indium atom obeys, for $\epsilon  < 3$, 
\begin{equation}
\label{v}
D_n(\epsilon) \sim \frac{3 (1 + \epsilon)}{\pi (3 \pi \epsilon - 7 \epsilon + 3 \pi - 3)} \; \frac{\ln(n)}{n},
\end{equation}
and $D_n(\epsilon = 3) = 0$. Hence, $D_n(\epsilon)$ is a monotoneously growing function of $\epsilon$ for
 $\epsilon  < 3$, and is discontinuous at $\epsilon = 3$.  The ratio $D_n(\epsilon = \epsilon^*)/D_n(\epsilon = 0)$
is now of order of $1.9$, which shows that also for infinite lattices 
asymmetric exchange 
probabilities enhance the ${\rm In}$ diffusivity.

Finally, similarly to the approach
 of Ref.\cite{hilhorst1}, we derive from Eqs.(\ref{4}) and (\ref{P}), 
and the asymptotical expansions in Eqs.(\ref{N}) to (\ref{K}),  
the asymptotic form of the
probability distribution $P_n({\bf X})$. We find that as $n \to \infty$,  the distribution function of 
two scaling variables
\begin{equation}
\label{j}
\eta_{1,2} = \Big(\frac{4 \pi [(3 \pi - 7) \epsilon + 3 (\pi -1)]}{3 (1 + \epsilon)   \ln(n)} \Big)^{1/2} \; x_{1,2},
\end{equation}
converges to a limiting, non-Gaussian form:
\begin{equation}
\label{jj}
P(\eta_1, \eta_2) = \frac{1}{2 \pi} K_0\left(\sqrt{\eta_1^2 +  \eta_2^2}\right),
\end{equation}
where $K_0$ is the modified Bessel function of the zeroth order. 
Note that for $\epsilon = 0$, 
the results in Eqs.(\ref{G}), (\ref{j}) and (\ref{jj}), as well as $D_n(\epsilon)$
in Eqs.(\ref{diff}) and (\ref{v}),
coincide with the ones
obtained in Ref.\cite{hilhorst1}.

In conclusion, we have presented here 
an exact lattice solution of several versions of the model
originally devised by  van
Gastel et al \cite{saar1,saar2,saar3} 
to describe dynamics of an impure, Indium atom within the 
first layer of a Cu(001) surface. We have evaluated, 
for arbitrary values of the asymmetry parameter $\epsilon$, the
 long-time asymptotical behavior of the impure atom MSD, 
as well as the limiting scaling 
forms of the 
probability distribution.  
Our analytical results can be useful for further interpretation of the
 experimental data on dynamics of impure atoms in close-packed surface layers.

\end{document}